\documentclass[aip,pop,12pt,preprint,amsmath,amssymb,groupedaddress]{revtex4-1} 

\usepackage{bm}
\usepackage{graphicx}

\usepackage{mathptmx} 

\DeclareSymbolFont{newfont}{OML}{cmm}{m}{it}
\DeclareMathSymbol{\Varrho}{3}{newfont}{37}

\newcommand{\wt}{\widetilde}

\newcommand{\ave}[1]{{\left<#1\right>}}

\newcommand{\order}[1]{{\mathcal{O}\left(#1\right)}}

\newcommand{\rmd}{\text{d}}

\newcommand{\taud}{\ensuremath{\tau_\text{d}}}

\newcommand{\tauw}{\ensuremath{\tau_\text{w}}}

\newcommand{\Phiave}{\ensuremath{\ave{\Phi}}}
\newcommand{\Phirms}{\ensuremath{\Phi}_\text{rms}}

\newcommand{\Phitilde}{\ensuremath{\widetilde{\Phi}}}
\newcommand{\Phiwt}{\ensuremath{\widetilde{\Phi}}}

\newcommand{\Arms}{\ensuremath{A_\text{rms}}}
\newcommand{\aveA}{\ensuremath{\ave{A}}}
\newcommand{\Aave}{\ensuremath{\langle{A}\rangle}}

\newcommand{\Kave}{\ensuremath{\langle{K}\rangle}}

\newcommand{\Eqref}[1]{Eq.~\eqref{#1}}
\newcommand{\Eqsref}[1]{Eqs.~\eqref{#1}}
\newcommand{\Figref}[1]{Fig.~\ref{#1}}

\newcommand{\JNM}{\textit{J.~Nuclear Mater.}}

\newcommand{\NF}{\textit{Nucl.\ Fusion}}
\newcommand{\NME}{\textit{Nucl.\ Mater.\ Energy}}

\newcommand{\PFA}{\textit{Phys.\ Fluids~A}}

\newcommand{\PPCF}{\textit{Plasma Phys.\ Contr.\ Fusion}}
\newcommand{\PFR}{\textit{Plasma Fusion Res.}}
\newcommand{\PHP}{\textit{Phys.\ Plasmas}}
\newcommand{\PLA}{\textit{Phys.\ Lett.~A}}
\newcommand{\PP}{\textit{Phys.\ Plasmas}}
\newcommand{\PR}{\textit{Phys.\ Rev.}}
\newcommand{\PRA}{\textit{Phys.\ Rev.~A}}
\newcommand{\PRE}{\textit{Phys.\ Rev.~E}}
\newcommand{\PS}{\textit{Phys.\ Scripta}}

\newcommand{\BTSJ}{\textit{Bell Sys.\ Tech. J.}}

\newcommand{\JFM}{\textit{J.~Fluid Mech.}}
\newcommand{\PRL}{\textit{Phys.~Rev.\ Lett.}}

\begin{document}

\title{Intermittent fluctuations due to uncorrelated Lorentzian pulses}

\author{O.~E.~Garcia}
\email{odd.erik.garcia@uit.no}
\author{A.~Theodorsen}
\email{audun.theodorsen@uit.no}

\affiliation{Department of Physics and Technology, UiT The Arctic University of Norway, N-9037 Troms{\o}, Norway}

\date{\today}

\begin{abstract}
Fluctuations due to a super-position of uncorrelated Lorentzian pulses with a random distribution of amplitudes and duration times are considered. These are demonstrated to be strongly intermittent in the limit of weak pulse overlap, resulting in large skewness and flatness moments. The characteristic function and the lowest order moments are derived, revealing a parabolic relationship between the skewness and flatness moments. Numerical integration reveals the probability density functions in the case of exponential and Laplace distributed pulse amplitudes. This stochastic model describes the intermittent fluctuations and probability densities with exponential tails commonly observed in turbulent fluids and magnetized plasmas.
\end{abstract}

\maketitle

From numerous experiments on and model simulations of fluids and magnetized plasmas it has been demonstrated that chaotic fluctuations have an exponential frequency power spectral density.\cite{mm-pre,mm-prl,atten,libchaber,brandstater,streett,mensour,paul,mckee,hornung,pace-prl,pace-php,mm-ppcf,mrm-ppcf,zhu} In many cases this has been associated with Lorentzian pulses in the underlying time series.\cite{mm-pre,mm-prl,hornung,pace-prl,pace-php,mm-ppcf,mrm-ppcf,zhu} Recently, a novel analysis method was applied in order to separate the complexity and randomness of the fluctuations.\cite{mm-ppcf,mrm-ppcf,zhu} However, intermittency of the fluctuations and the probability density function has usually not been investigated. This is despite the fact that in many turbulent fluid and plasma systems it has been found that the fluctuations are strongly intermittent and that there is an exponential tail in the probability density function for large fluctuation amplitudes.\cite{heslot,castaing,sano,deluca,massaioli,christie,takeshita,julien,garcia-ppcf,garcia-pre,antar1,antar2,garcia-esel1,garcia-esel2,garcia-esel3,graves,labombard,garcia-tcv1,garcia-tcv2,militello1,militello2,garcia-acm,garcia-psi,theodorsen-tcv,garcia-kstar,ghp}

Here a stochastic model is presented which describes all these features of the fluctuations by describing them as a super-position of uncorrelated Lorentzian pulses with a random distribution amplitudes and duration times. General expressions for the lowest order moments and the characteristic function are derived. The fluctuations are shown to be strongly intermittent when the ratio of the average pulse duration and waiting times is small, most clearly manifested by large skewness and flatness moments. In the opposite limit with significant pulse overlap, the probability density function approaches a normal distribution. There is a universal parabolic relationship between the skewness and flatness moments. A closed form of the characteristic function for the process is derived for exponential and Laplace distributed pulse amplitudes. The corresponding probability density functions are calculated numerically and shown to have exponential tails for large fluctuation amplitudes in the case of weak pulse overlap. The results presented here complement previous work on the same stochastic process with emphasis on the frequency power spectral density presented in Ref.~\onlinecite{gt-l} and \onlinecite{gt-lpsd}.

Consider a stochastic process given by a super-position of $K$ uncorrelated pulses with a fixed shape in a time interval of duration $T$,\cite{gt-l,gt-lpsd,garcia-prl,garcia-php,garcia-psd,rice1,kube-php,theodorsen-php,theodorsen-ps}
\begin{equation}\label{shotnoise}
\Phi_K(t) = \sum_{k=1}^{K(T)} A_k\phi\left( \frac{t-t_k}{\tau_k} \right) ,
\end{equation}
where each pulse labeled $k$ is characterized by an amplitude $A_k$, arrival time $t_k$, and duration $\tau_k$, all assumed to be uncorrelated and each of them independent and identically distributed. The number of pulses $K$ in an interval of duration $T$ is given by the Poisson distribution,
\begin{equation} \label{poisson}
P_K(K|T) = \frac{1}{K!}\left(\frac{T}{\tauw}\right)^K\exp\left(-\frac{T}{\tauw} \right) ,
\end{equation}
with mean value
\begin{equation}
\Kave = \sum_{K=0}^{\infty} KP_K(K|T) = \frac{T}{\tauw} .
\end{equation}
Here and in the following, angular brackets denote the average of the argument over all random variables unless otherwise explicitly stated. From this it follows that the waiting times between the pulses are exponentially distributed with mean value $\tauw$ and that the pulse arrival times are uniformly distributed on the time interval under consideration, that is, their probability density function is given by $1/T$.

The pulse duration times $\tau_k$ are assumed to be randomly distributed with probability density $P_\tau(\tau)$, and the average pulse duration time is defined by
\begin{equation}\label{Ptaud}
\taud = \langle \tau \rangle = \int_0^\infty \rmd\tau\,\tau P_\tau(\tau) .
\end{equation}
The pulse shape $\phi(\theta)$ is taken to be the same for all events in \Eqref{shotnoise} and is in this study given by the Lorentzian pulse
\begin{equation} \label{lorentzian}
\phi(\theta) = \frac{1}{\pi}\frac{1}{1+\theta^2} .
\end{equation}
The integral of the $n$-th power of the Lorentzian pulse shape is given by
\begin{equation} \label{pulseint}
I_n = \int_{-\infty}^{\infty} \rmd\theta\,\left[ \phi(\theta) \right]^n = \frac{1}{\pi^{n-1/2}}\frac{\Gamma(n-1/2)}{\Gamma(n)} ,
\end{equation}
where $\Gamma$ is the Gamma function. The lowest order pulse function integrals are given by $I_1=1$, $I_2=1/2\pi$, $I_3=3/8\pi^2$ and $I_4=5/16\pi^3$.

Starting with the case of exactly $K$ events in a time interval of duration $T$, the mean value of the process is given by integrating over all random variables and neglecting end effects by taking the integration limits for the pulse arrival times $t_k$ to infinity, giving $\langle{\Phi_K}\rangle=\taud I_1\ave{A}K/T$. Taking into account that the number of pulses $K$ is also a random variable and averaging over this as well gives the mean value for the stationary process,
\begin{equation} \label{phiave}
\ave{\Phi} = \frac{\taud}{\tauw}\,\ave{A} .
\end{equation}
The mean value is large when there is significant overlap of pulse events, that is, for long pulse durations and short pulse waiting times.

\begin{figure}
\includegraphics[width=8.5cm]{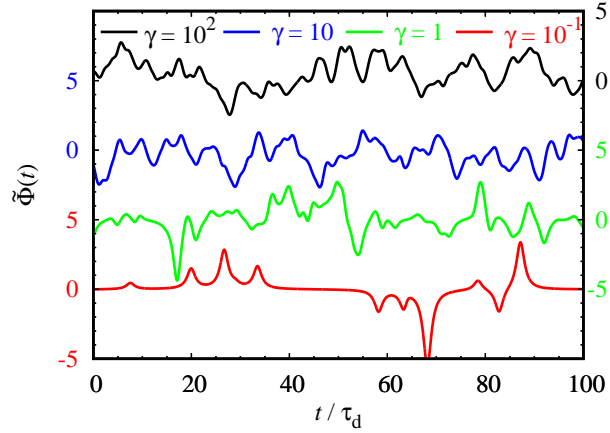}
\caption{Realizations of the stochastic process for Lorentzian pulses with constant duration $\taud$ and Laplace distributed pulse amplitudes. The degree of pulse overlap is determined by the intermittency parameter $\gamma=\taud/\tauw$.}
\label{fig:Phiraw}
\end{figure}

The variance can similarly be calculated by averaging the square of the random variable, giving $\langle{\Phi^2}\rangle=\Phiave^2+\taud I_2\langle{A^2}\rangle/\tauw$. The square of the root mean square (rms) value is therefore
\begin{equation} \label{phivariance}
\Phirms^2 = \frac{1}{2\pi}\frac{\taud}{\tauw}\,\langle{A^2}\rangle .
\end{equation}
For reasons to become clear presently, the ratio of the average pulse duration and waiting times,
\begin{equation}
\gamma = \frac{\taud}{\tauw} ,
\end{equation}
is referred to as the \emph{intermittency parameter} of the process. In the case of a finite mean value, the relative fluctuation level for Lorentzian pulses is given by
\begin{equation}
\frac{\Phirms^2}{\Phiave^2} = \frac{1}{2\pi\gamma}\frac{\langle{A^2}\rangle}{\Aave^2} ,
\end{equation}
which is large when there is weak overlap of the pulse structures. This is clearly illustrated in \Figref{fig:Phiraw}, which shows realizations of the process for Lorentzian pulses with constant duration and a Laplace distribution of the pulse amplitudes. Here the rescaled variable with zero mean and unit standard deviation
\begin{equation}\label{Phitilde}
\Phiwt = \frac{\Phi-\Phiave}{\Phirms} ,
\end{equation}
has been introduced. It is noted that the condition $\Phi>0$ implies $\Phitilde>-\Phiave/\Phirms$. For large values of $\gamma$ there is significant overlap of pulse structures. This results in a small relative fluctuation level and realizations of the process resemble random noise. For small values of $\gamma$, the time series are dominated by large-amplitude bursts and the process is strongly intermittent with large relative fluctuations. The intermittency is quantified by the skewness and flatness moments, which follow from the characteristic function for the process.

The characteristic function $C_\Phi(u)$ for a random variable is the Fourier transform of the probability density function $P_\Phi(\Phi)$, defined by
\begin{equation}
C_\Phi(u) = \int_{-\infty}^{\infty} \rmd\Phi\,P_\Phi(\Phi)\exp(i\Phi u) .
\end{equation}
The characteristic function for a sum of independent random variables is the product of their individual characteristic functions. The conditional probability density function for exactly $K$ uncorrelated pulses in a time interval of duration $T$ is
\begin{equation}
P_\Phi(\Phi | K) = \frac{1}{2\pi}\int_{-\infty}^{\infty} \rmd u\,\exp(-i\Phi u) \langle{\exp(iA_k\phi_k u)}\rangle^K ,
\end{equation}
where the characteristic function for each pulse $\phi_k=\phi((t-t_k)/\tau_k)$ is
\begin{equation} \label{charfunc}
\langle{\exp(iA_k\phi_k u)}\rangle = \int_{-\infty}^{\infty} \rmd A_k\,P_A(A_k) \int_0^\infty \rmd\tau_k\,P_\tau(\tau_k) \int_0^T \frac{\rmd t_k}{T}\,\exp\left[ iuA_k\phi\left( \frac{t-t_k}{\tau_k} \right)\right] .
\end{equation}
The probability density function for the random variable $\Phi$ is thus
\begin{equation}
P_\Phi(\Phi) = \sum_{K=0}^{\infty} P_\Phi(\Phi|K) P_K(K|T)
= \frac{1}{2\pi}\int_{-\infty}^{\infty} \rmd u\,\exp\left( - i\Phi u + \frac{T}{\tauw}\,\langle\exp(iA_k\phi_k u)\rangle - \frac{T}{\tauw} \right) ,
\end{equation}
where $P_K(K|T)$ is the Poisson distribution given by \Eqref{poisson}. The stationary probability density function for $\Phi$ is obtained by extending the integration limits for $t_k$ to infinity and making the change of integration variable given by $\theta=(t-t_k)/\tau_k$ in \Eqref{charfunc}. This leads to the desired result,
\begin{equation}\label{CPhiu}
C_\Phi(u) = \exp\left( \gamma \int_{-\infty}^{\infty} \rmd A\,P_A(A) \int_{-\infty}^{\infty} \rmd \theta\,\left[ \exp(iuA\phi(\theta))-1 \right] \right) ,
\end{equation}
which notably is independent of the distribution function for the pulse duration times. The characteristic function for the stationary process is determined by the pulse shape, the amplitude distribution and the degree of pulse overlap.

By expanding the exponential function in \Eqref{CPhiu} and then performing the integration over $\theta$, the logarithm of the characteristic function for the process is
\begin{equation} \label{lncf2}
\ln{C_\Phi(u)} = \sum_{n=1}^\infty \gamma I_n\langle{A^n}\rangle \,\frac{(iu)^n}{n!} ,
\end{equation}
where $I_n$ is defined by \Eqref{pulseint}. The cumulants $\kappa_n$ are the coefficients in the expansion of the logarithm of the characteristic function for $P_\Phi$. For the stochastic process considered here, the cumulants are thus given by 
\begin{equation} \label{cumulant}
\kappa_n = \gamma I_n \ave{A^n} .
\end{equation} 
From the cumulants, the lowest order moments are readily obtained. A formal power series expansion shows that the characteristic function is related to the raw moments of $\Phi$,
\begin{equation} \label{cfs}
C_\Phi(u) = 1 + \sum_{n=1}^{\infty} \ave{\Phi^n}\,\frac{(iu)^n}{n!} .
\end{equation}
Further expanding the logarithmic function in \Eqref{lncf2} and using \Eqref{cfs}, 
it follows that the lowest order centred moments $\mu_n=\ave{(\Phi-\ave{\Phi})^n}$ are related to the cumulants by the relations $\mu_2=\kappa_2$, $\mu_3=\kappa_3$ and $\mu_4=\kappa_4+3\kappa_2^2$. From this, general expressions for the skewness and flatness moments are readily obtained,\cite{garcia-prl,garcia-php}
\begin{subequations} \label{SandF-L}
\begin{align}
S_\Phi & = \frac{3}{4} \left( \frac{2}{\pi\gamma} \right)^{1/2} \frac{\langle{A^3}\rangle}{\langle{A^2}\rangle^{3/2}} ,
\\
F_\Phi & = 3 + \frac{5}{4\pi\gamma}\frac{\langle{A^4}\rangle}{\langle{A^2}\rangle^2} .
\end{align}
\end{subequations}
Both these moments increase with decreasing $\gamma$, clearly demonstrating the intrinsic intermittent features of a process composed by a super-position of uncorrelated pulses. For a symmetric amplitude distribution the skewness moment vanishes together with the mean value of the random variable. More generally, \Eqsref{SandF-L} imply that there is a parabolic relationship between the skewness and flatness moments,\cite{garcia-prl}
\begin{equation} \label{parabolic}
F_\Phi = 3 + \frac{2\pi^2}{5}\frac{\ave{A^2}\ave{A^4}}{\ave{A^3}^2}\,S_\Phi^2 .
\end{equation}
This relation holds for any amplitude and duration time distributions as far as the amplitude moments exist.

The results presented above show that the skewness and excess flatness moments vanish in the limit $\gamma\rightarrow\infty$. It can be demonstrated that the probability density function for $\Phiwt$ then approaches a normal distribution, independent of the details of the pulse shape and amplitude and duration time distributions. The stationary distribution $P_\Phi$ can be written in terms of the characteristic function given by \Eqref{cfs},
\begin{equation}
P_\Phi(\Phi) = \frac{1}{2\pi}\int_{-\infty}^{\infty} \rmd u\,\exp\left( - i\Phi u + \sum_{n=1}^\infty \frac{\kappa_n(iu)^n}{n!} \right) ,
\end{equation}
where the cumulants are given by \Eqref{cumulant}. In the limit of large $\gamma$ the exponential function can be expanded as a power series in $u$. Integrating term by term then gives\cite{rice1,garcia-prl,garcia-php}
\begin{equation} \label{Gausslimit}
\lim_{\gamma\rightarrow\infty} P_{\wt{\Phi}}(\wt{\Phi}) = \lim_{\gamma\rightarrow\infty}
\frac{1}{(2\pi)^{1/2}}\,\exp\left( -\frac{\wt{\Phi}^2}{2} \right) \left[ 1 + \frac{\mu_3}{3!\Phirms^3(2\pi)^{1/2}}( \wt{\Phi}^3 - 3\wt{\Phi} ) + \order{1/\gamma} \right] .
\end{equation}
The terms inside the square bracket in \Eqref{Gausslimit} are of order $1$, $1/\gamma^{1/2}$ and $1/\gamma$, respectively. The last of these represents the sum of the remaining terms in the expansion. This shows how the probability density function for $\Phitilde$ approaches a normal distribution in the limit of large $\gamma$. The transition to normal distributed fluctuations is expected from the central limit theorem, since in this case a large number of uncorrelated pulses contribute to $\Phiwt$ at any given time. The normal limit is valid for arbitrary pulse shapes and amplitude and duration time distributions as far as the cumulants are finite.

By introducing the rescaled variable $\Phitilde$ defined by \Eqref{Phitilde}, it is straight forward to show that the corresponding characteristic function is given by
\begin{equation}
C_{\wt{\Phi}}(v) = \exp\left( - i\,\frac{\Phiave}{\Phirms}\,v \right) C_\Phi\left( \frac{v}{\Phirms} \right) ,
\end{equation}
where $C_\Phi(u)$ is given by \Eqref{lncf2}. Closed analytical expressions for $C_{\wt{\Phi}}$ will be obtained for two relevant amplitude distributions. Consider first the case of an exponential distribution of the pulse amplitudes,
\begin{equation}
P_A(A) = \frac{1}{\aveA}\,\exp\left( - \frac{A}{\aveA} \right) ,
\end{equation}
where $\aveA$ is the mean pulse amplitude and $P_A$ is defined only for positive amplitudes, $A>0$. In this case, the raw amplitude moments are given by $\langle{A^n}\rangle=n!\aveA^n$. For the stationary process it follows that the mean value is finite, $\Phiave=\gamma\aveA$, and the variance is given by $\Phirms^2=\gamma\aveA^2/\pi$, giving the relative fluctuation level $\Phirms/\Phiave=1/(\pi\gamma)^{1/2}$. The skewness and flatness moments become $S_\Phi=9/4(\pi\gamma)^{1/2}$ and $F_\Phi=3+15/2\pi\gamma$, respectively. Note that in this case there is a parabolic relationship between the skewness and flatness moments given by $F_\Phi=3+40S_\Phi^2/27$. For positive definite pulse amplitudes, the condition $\Phi>0$ corresponds to $\Phitilde>-(\pi\gamma)^{1/2}$.

\begin{figure}
\includegraphics[width=8.5cm]{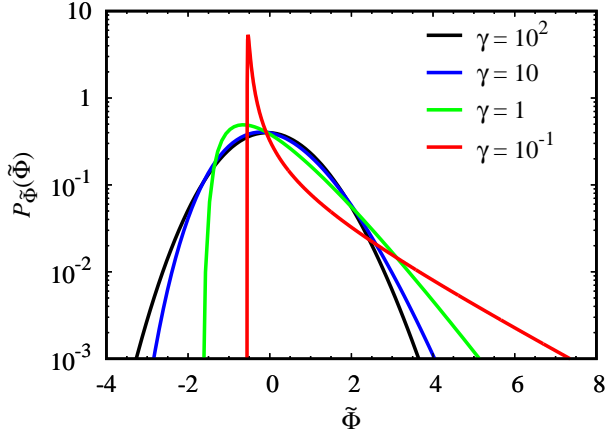}
\caption{Probability density functions for a super-position of uncorrelated Lorentzian pulses with an exponential amplitude distribution and various values of the intermittency parameter $\gamma$.}
\label{fig:pdf-exp}
\end{figure}

The characteristic function for an exponential amplitude distribution is given by
\begin{equation}
\ln{C_\Phi(u)} = \pi\gamma\,\frac{i\aveA u}{(\pi-i\aveA u)^{1/2}} .
\end{equation}
Making the substitution $w=v/(\pi\gamma)^{1/2}$, the characteristic function for the rescaled variable $\Phitilde$ is given by
\begin{equation}
P_{\wt{\Phi}}(\Phitilde) = \frac{1}{2}\left(\frac{\pi}{\gamma}\right)^{1/2} \int_{-\infty}^{\infty} \rmd w\,\exp\left( -i\sqrt{\pi\gamma}\,(\Phitilde+\sqrt{\pi\gamma})w + \frac{i\pi\gamma w}{\sqrt{1-iw}} \right) . 
\end{equation}
It is noted that for any function $f(w)$ with the property $f^{\dag}(w)=f(-w)$, where the dagger denotes the complex conjugate, the following relation holds,
\begin{equation}\label{ccreal}
\int_{-\infty}^{\infty} \rmd w\,f(w) = 2\int_0^\infty \rmd w\,\mathcal{R}[f(w)] ,
\end{equation}
where $\mathcal{R}$ denotes the real part of the argument. Any characteristic function satisfies this condition, so the probability density can be written as an integral over a function that takes only real values,
\begin{multline}\label{PPhiwt-exp}
P_{\wt{\Phi}}(\Phitilde) = \left(\frac{\pi}{\gamma}\right)^{1/2} \int_0^\infty \rmd w\,\exp\left( - \frac{\gamma\pi w\sin\left(\frac{1}{2}\,\arctan{w}\right)}{(1+w^2)^{1/4}} \right)
\\
\times \cos\left( \pi\gamma w + \sqrt{\pi\gamma}\,\Phitilde w - \frac{\pi\gamma w\cos\left(\frac{1}{2}\,\arctan{w}\right)}{(1+w^2)^{1/4}} \right) .
\end{multline}
This expression is suitable for numerical integration and the distribution function for the rescaled variable is presented in \Figref{fig:pdf-exp} for various values of the intermittency parameter. The probability density function is unimodal for all values of the intermittency parameter and has an exponential tail towards large fluctuation amplitudes for small values of $\gamma$. For large values of $\gamma$, the probability density function for $\Phitilde$ approaches a normal distribution with vanishing mean and unit standard deviation.\cite{rice1,garcia-prl,garcia-php}

\begin{figure}
\includegraphics[width=8.5cm]{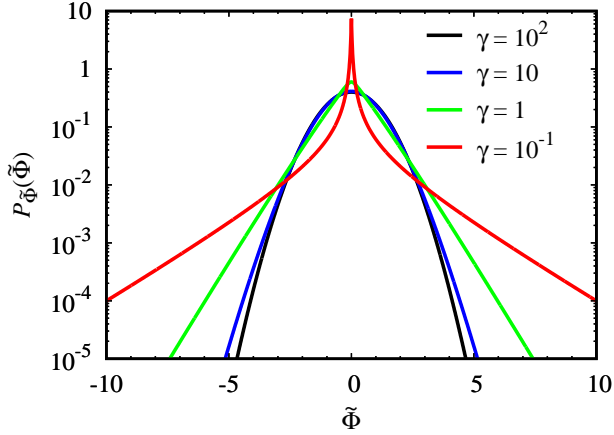}
\caption{Probability density functions for a super-position of uncorrelated Lorentzian pulses with a Laplace amplitude distribution and various values of the intermittency parameter $\gamma$.}
\label{fig:pdf-laplace}
\end{figure}

Allowing both positive and negative pulse amplitudes, the symmetric Laplace distribution with vanishing mean is of particular interest,
\begin{equation}
P_A(A) = \frac{1}{2^{1/2}\Arms}\,\exp\left( - \frac{2^{1/2}|A|}{\Arms} \right) ,
\end{equation}
where $\Arms$ is the standard deviation, $\langle{A^2}\rangle=\Arms^2$. The odd moments for this distribution vanish, while the even moments are given by $\langle{A^{2n}}\rangle=(2n)!(\Arms/2^{1/2})^{2n}$ for positive integers $n$. For this symmetric distribution, both the mean value and the skewness moment of the random variable vanish, $\Phiave=0$ and $S_\Phi=0$. The variance of the random variable is now given by $\Phirms^2=\gamma\Arms^2/2\pi$, while the flatness moment is $F_\Phi=3+15/2\pi\gamma$. The latter is the same as for exponentially distributed amplitudes discussed above. Moreover, the characteristic function can be expressed in closed form,
\begin{equation}
\ln{C_\Phi(u)} = - \frac{i\sqrt{\pi}\gamma\Arms u}{2} \frac{( 2\pi-i\sqrt{2}\Arms u )^{1/2}-( 2\pi + i\sqrt{2}\Arms u )^{1/2}}{\left( 4\pi^2 + 2\Arms^2 u^2 \right)^{1/2}} .
\end{equation}
Again using the relation given by \Eqref{ccreal} and the change of integration variable defined by $w=v/(\pi\gamma)^{1/2}$, an expression for the probability density function that is suitable for numerical integration is obtained,
\begin{equation}
P_{\wt{\Phi}}(\Phitilde) = \left(\frac{\gamma}{\pi}\right)^{1/2} \int_0^\infty \rmd w\,\exp\left( - \frac{\pi\gamma w\sin\left( \frac{1}{2}\,\arctan{w} \right)}{2(1+w^2)^{1/4}} \right)\cos\left( \sqrt{\pi\gamma}\,\Phitilde w \right) .
\end{equation} 
This distribution is presented in \Figref{fig:pdf-laplace} for various values of the intermittency parameter $\gamma$. For small values of $\gamma$, the distribution is strongly peaked and has exponential tails for large fluctuation amplitudes. In this case, the process spends long time intervals close to zero value between pulse arrivals, resulting in strong intermittency as shown in \Figref{fig:Phiraw}. In the limit $\gamma\rightarrow\infty$, the probability density function for $\Phitilde$ approaches a normal distribution with vanishing mean and unit standard deviation as discussed above.\cite{rice1,garcia-prl,garcia-php}

Intermittent fluctuations in chaotic and turbulent continuum systems have here been investigated by a stochastic model describing these as a super-position of uncorrelated Lorentzian pulses. The reference model has been extended to include a random distribution of pulse amplitudes and duration times. The intermittency of the system is determined by the degree of pulse overlap, quantified by the ratio of the average pulse duration and waiting times. When this parameter is large, many pulses contribute to the process at any given time, and the probability density function approaches a normal distribution as expected from the central limit theorem. In the opposite limit where pulses generally appear isolated, the process is strongly intermittent with large relative fluctuation level and skewness and flatness moments.

The characteristic function, and therefore the moments and probability density function, are not affected by a random distribution of the pulse duration times. The characteristic function can be calculated in closed form for several relevant pulse amplitude distributions. Numerical solutions for the probability density function has been obtained for exponentially and Laplace distributed pulse amplitudes. In both cases, there is an exponential tail for large fluctuation amplitudes in the strong intermittency limit. This is a well known feature of turbulent thermal convection and magnetized plasmas.

In summary, the stochastic model given by a super-position of uncorrelated Lorentzian pulses describes many of the salient features in chaotic and turbulent fluids and magnetized plasmas. This includes an exponential frequency spectrum in the case of constant pulse duration. Here the first predictions have been presented for the intermittency of the fluctuations and the probability density function, which has so far not been investigated in systems where Lorentzian pulses have been identified.\cite{mm-pre,mm-prl,hornung,pace-prl,pace-php,mm-ppcf,mrm-ppcf,zhu} On the other hand, it was recently established that the frequency power spectral density has an exponential shape for all values of the intermittency parameter in the case of constant pulse duration, thus being independent of the degree of pulse overlap.\cite{gt-l,gt-lpsd}


This work was supported with financial subvention from the Research Council of Norway under grant 240510/F20. The authors acknowledge the generous hospitality of the MIT Plasma Science and Fusion Center where this work was conducted.

\end{document}